\documentclass[useAMS,usenatbib]{mn2e}
\usepackage{graphicx}
\usepackage{rotating}
\usepackage{longtable}
\usepackage{lscape}
\def\lsim{\mathrel{\rlap{\lower 3pt \hbox{$\sim$}} \raise 2.0pt \hbox{$<$}}}
\def\gsim{\mathrel{\rlap{\lower 3pt \hbox{$\sim$}} \raise 2.0pt \hbox{$>$}}}

\def\Fermi{\emph{Fermi}}
\title[A new model for Extragalactic Gamma-ray Background]
{A new model for the extragalactic $\gamma$-ray background}
\author[Cavadini et al.]{
	M.~Cavadini$^{1}$\thanks{E-mail: mcavadini@dfm.uninsubria.it}, 
        R.~Salvaterra$^{1}$, and
       	F.~Haardt$^{1,2}$\\
       	$^{1}$ Dipartimento di Fisica e Matematica, Universit\`{a} degli Studi dell'Insubria, via Valleggio 11, I-22100 Como, Italy\\
	$^{2}$INFN, Sezione di Milano-Bicocca, I-20126 Milano, Italy
       	}
	
\begin{document}
\date{ }
\maketitle
\label{firstpage}

\begin{abstract}
We present a two-parameter model of the extragalactic $\gamma$-ray background (EGB) in the 
0.1-100 GeV range as measured by the Large Area Telescope (LAT) onboard the {\it Fermi} satellite.  
The EGB can be fully explained as the sum of three distinct components, namely blazars,
non-beamed AGNs (Seyfert galaxies and QSOs), and cosmic rays from star-forming galaxies. 
The contribution to the background from beamed sources is obtained by fitting 
the {\it Fermi}-LAT blazar differential number counts assuming that the $\gamma$-ray luminosity function 
is directly proportional to the radio luminosity function of FRI and FRII galaxies. The high energy emission from non-beamed AGNs is instead determined  
by popular synthesis models of the observed X-ray background. Finally, the EGB is fit by adding a third component 
arising from pion decay in cosmic rays, assuming that such component is closely linked to the cosmic star formation history. 
We find that blazars dominate at energies $\gsim$ 10 GeV, for $E\lsim 0.2$ GeV the main contribution is from
non-beamed AGNs, while cosmic rays are required in between. 
Because of absorption due to interaction of $\gamma$-rays with 
the extra-galactic background light, our model falls short at the highest energies probed by LAT, ($\gsim 70$ GeV), leaving 
room to a possible contribution from dark matter particle annihilation. As an example, 
a particle of mass $\simeq 0.5$ TeV and cross section $\langle \sigma v \rangle \simeq 5 \times  10^{-26}$ cm$^3$ s$^{-1}$ 
can accomodate the data. 

\end{abstract}



\section{Introduction}

The extragalactic $\gamma$-ray background (hereafter EGB) represents a
fascinating challenge since his first detection by {\emph {SAS 2}} satellite
above 30 MeV (Fichtel, Simpson, \& Thomson 1978). 
The Large Area Telescope (LAT) on board the \Fermi ~\emph{Gamma Ray Space Telescope}  has
provided new EGB data in the range 0.1-100 GeV, after one year of
observations (Abdo et al. 2010a).
The EGB is the sum of resolved extragalactic sources, and diffuse emission due to unresolved sources and/or 
intrinsically diffuse radiation, and its measure requires an accurate modeling of galactic foregrounds. In the 
following we will refer to the EGB as the sum of the diffuse
component  {\it and} resolved (by {\it FERMI}) sources. The EGB intensity above 100 MeV as measured by {\it Fermi}-LAT 
is $I_{\rm{EGB}}(>\rm{100 ~MeV})~=1.42 \times 10^{-5}$ cm$^{-2}$ s$^{-1}$ sr$^{-1}$, 
where resolved sources accounts for $\simeq 27$\% of the emission, the rest being ascribed to the diffuse component.

BL Lac objects and Flat Spectrum Radio Quasars (FSRQs), collectively known as blazars, are the most
common sources in the extragalactic $\gamma$-ray sky in the 0.1-100 GeV range. Therefore   
many authors considered these sources as the main contributors to the EGB (Stecker \& Salamon 1996;
M\"ucke \& Pohl 2000; Inoue \& Totani 2009; Abazajian, Blanchet, \& Harding 2010a; Neronov \& Semikoz 2011; 
for a review see Dermer 2007). 
Abdo et al. (2010b) recently showed that the contribution of blazars to the EGB is not dominant, 
and the focus has been shifted on different, potentially new, sources.  
As an example, Fields, Pavlidou, \& Prodanovi{\'c} (2010)
presented a model of the cosmic production of $\gamma$-rays powered by the 
interaction of cosmic rays with the interstellar medium in star-forming galaxies. The  model reproduces
quite well the behavior of the EGB up to 10 GeV, but it is not fully satisfactory at higher energies.

In the present {\it letter}, we propose a model of the EGB based on the $\gamma$-ray emission from  
blazars, non-beamed AGNs (Seyfert galaxies and QSOs), and star-forming galaxies.
We show that a combination of these three populations fully explains the EGB in the {\it Fermi}-LAT energy range. 

\section{Extragalactic $\gamma$-ray sources}
Our model employs three components, blazars, AGNs, and star-forming galaxies. First, we assume that 
the blazar $\gamma$-ray luminosity function (LF) is proportional to the radio LF of FRI and FRII 
galaxies. To convert radio to $\gamma$-ray luminosities we use the spectral energy distribution proposed by 
Fossati et al. (1998). AGNs are taken from population synthesis models of the X-ray background. For the star-forming 
galaxy emission we adopt a model of pion decay in cosmic rays. 

Our strategy is the following: we first fix the contribution 
of blazars to the EGB by fitting the {\it Fermi}-LAT differential number counts, having as sole fitting parameter the fraction 
of radio galaxies beamed toward the Earth, hence appearing as blazars. Then we add the fixed background from 
non-beamed AGNs (Inoue, Totani, \& Ueda 2008). Finally, the contribution of galaxies is obtained by fitting the EGB. Note that since the shape of the cosmic ray contribution is determined by pion decay and star formation history, the only free parameter is the star formation ``efficiency" of molecular hydrogen (Stecker \& Venters 2011). We checked that results do not change if we instead perform a simultaneous, 2-parameter fit to the number counts and to the background data. 

In the following subsections we briefly describe the details of our modeling. 
Throughout the paper we consider a concordance cosmology with 
H$_0=70$~km/s/Mpc, $\Omega_m = 0.3$, and
$\Omega_\Lambda=0.7$. 

\subsection{Blazars}\label{sec:blazar}

The blazar contribution (in photons s$^{-1}$ cm$^{-2}$ sr$^{-1}$ MeV$^{-1}$) to the EGB at the observed energy $E_0$ is 
\begin{eqnarray}\label{eq.blazars}
I_{\rm{blaz}}(E_{0}) &=& \frac{1}{4\pi} 
\int_0^{\infty} dz\, \frac{dV}{dz}
\int_{\log L_{\gamma}^{\mathrm{min}}}^{\log L_{\gamma}^{\mathrm{max}}} d\log L_\gamma\,
\frac{d\Phi_\gamma (L_\gamma,z)}{d\log L_\gamma} \nonumber\\
&\times&\frac{dn(L_\gamma,z)}{dE} e^{-\tau_{\gamma \gamma}(E_0,z)},
\end{eqnarray}
where $d\Phi_\gamma(L_\gamma,z)/d\log L_\gamma$ is the $\gamma$-ray LF and 
$L_{\gamma}$ is  $\nu L_\nu$ (in erg/s) at 100 MeV, 
$dn(L_\gamma,z)/dE$ is the unabsorbed photon flux per unit energy $E=E_0(1+z)$
measured on Earth of a blazar with luminosity $L_{\gamma}$ at redshift $z$, and $\tau_{\gamma \gamma}(E_{0},
z)$ is the optical depth for $\gamma-\gamma$ absorption. We adopt the extragalactic background light (EBL) model by Finke, Razzaque, \& Dermer (2010). 
In the above equation $dV/dz$ is the comoving cosmological volume. 
We set $\log L_{\gamma}^{\mathrm{min}}=43$ and $\log L_{\gamma}^{\mathrm{max}}=50$.

The number of sources $N(>F_{\rm ph})$ per steradian with photon flux
greater than $F_{\rm ph}$ is
\begin{equation}
N(>F_{\rm ph})= \frac{1}{4\pi}
  \int_{0}^{\infty}dz\, \frac{dV}{dz} 
 \int_{\log L_{\gamma}^{\mathrm{min}}}^{\log L_{\gamma}^{\mathrm{max}}} d\log L_\gamma\,
 \frac{d\Phi_\gamma(L_\gamma,z)}{d\log L_\gamma}.
\label{eq:counts}
\end{equation} 
The $\gamma$-ray LF of blazars is presently uncertain (for an estimate see, e.g., Abdo et al. 2009), 
so that one has to rely on the LFs computed in other bands, e.g., X-rays (Narumoto \& Totani 2006;
Inoue \& Totani 2009; Abazajian et al. 2010a), or radio (Draper \& Ballantyne 2009). We adopt here the 
radio LF at 151 MHz of FRI and FRII (Willot et al. 2001), assumed to be the parent populations of blazars: 
\begin{equation}\label{blaz}
\frac{\Phi_{\gamma}(L_{\gamma},z)} {d\log L_{\gamma}}=\kappa~\frac{\Phi_{R}(L_{R},z)}{
d\log L_{R}},
\end{equation}
where $L_{\rm R}$ is $\nu L_\nu$ at 151 MHz, and 
the constant $\kappa$ is the fraction of blazars over all radio galaxies, and it is out first fit parameter. 
In order to convert radio into 
$\gamma$-ray luminosity, we must rely on the blazar spectral energy distribution (SED). 
According to the SSC model, the blazar SED shows two broad peaks, the first located in the
IR/X-ray band, due to synchrotron emission from the relativistic jet, and the second in the $\gamma$-ray band, 
due to the inverse Compton scattering of synchrotron photons (e.g., Ghisellini et al. 1998). 
As pointed out by Fossati et al. (1998), a relation between
the radio and/or bolometric luminosity and the energy of the two peaks exists 
(the so-called {\it blazar sequence}, see also Donato et al. 2001). 
We use the SEDs computed by Inoue \& Totani (2009)  based
on the empirical determinations of Donato et al. (2001).

\subsection{Non-beamed AGNs}\label{sec:AGN}
Hot thermal electrons in accretion disk coronae of Seyfert galaxies and QSOs scatter off UV photons into 
the hard X-ray/$\gamma$-ray bands (e.g., Haardt \& Maraschi 1991, 1993). Such emission, filtered by absorbing 
material at a distance of several parsecs form the central source, is though to be responsible of the observed 
X-ray background (e.g., Setti \& Woltjer 1989; Madau, Ghisellini, \& Fabian 1994; Gilli, Comastri, \& Hasinger 2007). 
According to Inoue et al. (2008) 
the AGNs responsible of the hard X-ray background can also account for the 1-10 MeV EGB, provided that 
an extra non-thermal electron population is considered. As in Abazajian et al. (2010a), we use the model of Inoue et al. 2008 
with a powerlaw index of 
non-thermal electrons $=3.5$, a lower limit of the Lorentz factor
distribution $=4.4$.

\subsection{Star-forming Galaxies} \label{sec:galaxy}
The $\gamma$-ray spectrum of a star-forming galaxy is based on the assumption that $\gamma$-ray emission 
is due to the decay of $\pi^{0}$ mesons. The $\pi^{0}$ mesons form in the inelastic
collision between cosmic rays and the ISM. According to Stecker \& Venters (2011), 
the specific $\gamma$-ray photon spectrum $L_{\rm ph}$ (photons
s$^{-1}$ MeV$^{-1}$) of a star-forming galaxy 
is related to the average pionic $\gamma$-ray production spectrum per hydrogen atom 
$\langle q_{\rm{H}}(E_{0})\rangle$ (Dermer 1986; Mori 1997) as,
\begin{equation}
L_{\rm ph}(E_{0}) = \langle q_{\rm{H}}(E_{0})\rangle N_{\rm H},	
\end{equation}
where $N_{\rm H}$ is the total number of hydrogen atoms in the galaxy, both in atomic and molecular form. 

We adopt the \emph{Strong Coupling $\gamma$-ray - Star Formation Rate
Model} of Stecker \& Venters (2011), where $N_{\rm H}$ is related to
the star formation rate. According to the model, the star-forming galaxy contribution to the 
$\gamma$-ray background is
\begin{equation}
I_{\rm gal}(E_0)=\frac{1}{4\pi}\int_0^{\infty}{dz\frac{dl}{dz}\, 
\frac{(1+{\cal R})}{m_{\rm H} \xi(H_2)}\langle q_{\rm H}(E)\rangle\, \dot\rho_{\rm SFR}(z) e^{-\tau_{\gamma \gamma}(E_0,z)}},
\end{equation}
where $dl/dz=cH^{-1}(1+z)^{-1}$ with $H(z)$ the Hubble parameter,  
${\cal R}\sim 0.9$ is the ratio of atomic-to-molecular hydrogen density in star-forming galaxies (see Leroy et al. 2008), 
and $\dot\rho_{\rm SFR}$ is the cosmic star formation rate (we use the fit proposed by Li 2008). 
The parameter $\xi(H_2)$ (the star formation efficiency of 
molecular hydrogen, see Bigiel et al. 2008, Gnedin et al. 2009; Bauremeister et al. 2010) is the ratio between 
$\dot\rho_{\rm SFR}$ and the cosmic density of molecular hydrogen, and it is our second fit parameter. 
\begin{figure}
\begin{center}
\includegraphics[scale=0.4]{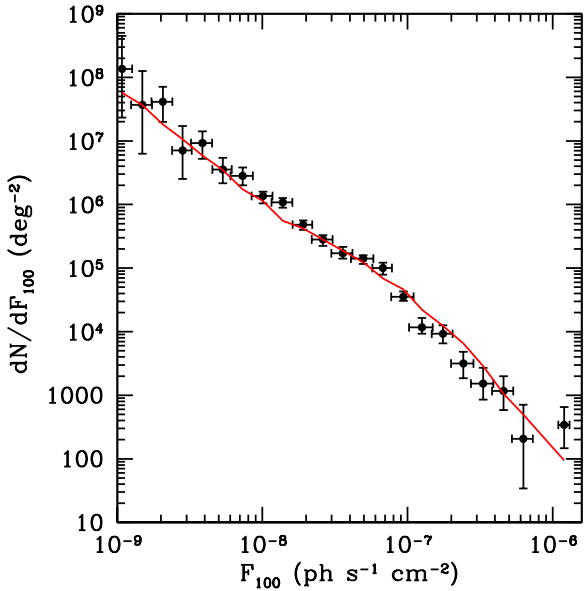}
\end{center}
\caption{The data points are the differential blazar number counts detected by {\it Fermi}-LAT as a function 
of the integrated flux above 100 MeV $F_{100}$ (Abdo et al 2010b). The solid red line is our best fit.} 
\label{fig:conteggi}
\end{figure}

\begin{figure}
\begin{center}
\includegraphics[scale=0.4]{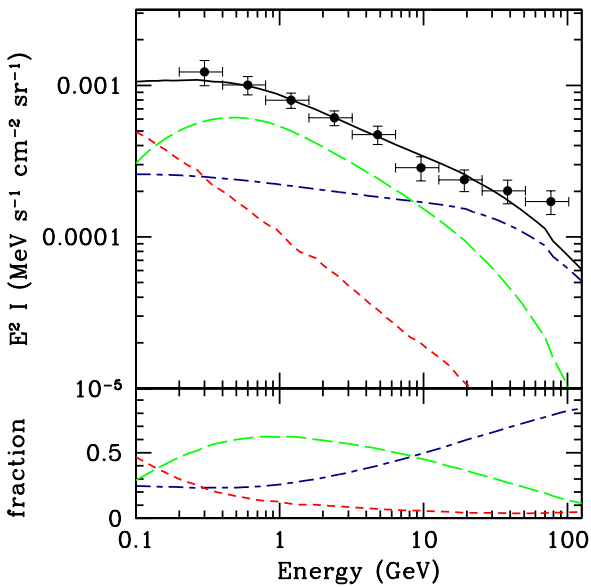}
\end{center}
\caption{
{\it Upper panel}: 
the EGB measured by {\it Fermi}-LAT (data points by Abdo et al. 2010a), and our fit to it.
The total emission ({\it solid black line}) is shown together with the contribution from blazars ({\it  long-short dashed blue line}), 
non-beamed AGNs ({\it short dashed red line}), and $\gamma$-ray emission from cosmic rays in 
star-forming galaxies ({\it long dashed green line}). {\it Lower panel}: fractional contributions of the three different 
components to the EGB.
}
\label{fig:fondo}
\end{figure}

\section{Results}
We use the model discussed in the previous section to fit the {\it Fermi}-LAT EGB. As already 
discussed, we first compute the contribution 
of blazars by fitting their {\it Fermi}-LAT differential number counts (Abdo et al. 2010b), then we add the background from 
non-beamed AGNs (Inoue et al. 2008), and finally we compute the contribution of galaxies by fitting the EGB. 
We obtain the same best fit values (within the errors) if we 
perform a simultaneous fit to the number counts of blazars {\it and} 
to the background. Results and numbers reported here refer to this second approach. 
We remark that the shape of the star-forming galaxies and blazar
components are determined a priori, while only their normalizations are allowed to vary. The 
contribution of non-beamed AGNs is instead totally fixed.

In fig.~\ref{fig:conteggi} we show our best fit to the differential blazar number counts, while
the best fit to the $\gamma$-ray background is shown in fig.~\ref{fig:fondo}, upper panel. In the lower panel of 
fig.~\ref{fig:fondo} we show the relative fractional contributions of the three different components.

The best fit parameter values we obtain are $\kappa=3.93 \pm 0.01 \times 10^{-4}$ 
and $\xi(H_2)=4.07 \pm 0.4 \times 10^{-10} ~\rm{yr^{-1}}$, with $\chi^{2}/\rm{d.o.f}=1.15$ for 29 d.o.f.. 
The best fit value of $\xi$, which represents the star formation efficiency of molecular hydrogen is in 
agreement with observational values  ($\xi \sim (5.25 \pm 2.5) \times
10^{-10} \rm{yr^{-1}}$, Leroy et al. 2008). The number ratio of blazars to radio galaxies $\kappa$ can be thought as a measure of the 
beaming factor of the relativistic jet, which in turn is related to the bulk Lorentz factor $\Gamma$. From $\kappa \sim 1/2\Gamma^2$ we derive $\Gamma \sim 35$.

\section{Discussion and Conclusions}\label{sec:discussion}

We proposed a two parameter model for the extragalactic $\gamma$-ray background (EGB) measured by {\it Fermi}-LAT. 
We showed that the EGB can be explained by a combination of blazars, non-beamed AGNs 
(Seyfert galaxies and QSOs), and high energy photons originating by the interaction of cosmic rays and ISM in star-forming galaxies.  

Our EGB model differs with respect to other existing similar studies (e.g., Inoue \& Totani 2009; Abazajian et al. 2010a; 
Stecker \& Venters 2011) in the $\gamma$-ray LF of blazars that we derive directly from the radio LF 
allowing only a free relative normalization (i.e., the LF faint and bright ends are fixed before fitting), and/or in the use of the 
Fossati et al. (1998) SED instead of a simple power law parametrization of blazar high energy spectrum.   

We found that non-beamed AGNs dominate the EGB below 0.2 GeV, while blazars take over above 10 GeV. Cosmic ray emission is then needed at intermediate energies at a level $\simeq 50\%$ of the total. Our model parameters are the relative ratio of blazars to radio galaxies, and the absolute normalization of the 
contribution from star-forming galaxies. The best fit value of the relative number of blazars with respect to radio galaxies can be translated into a bulk Lorentz factor of the 
relativistic jet $\Gamma \sim 35$, larger than the average value $\Gamma\sim 15$ estimated by Ghisellini et al. (2010). The two values could be reconciled if blazars commonly show secular $\gamma$-ray large variability which modulates the 1-year average flux, as recently proposed by Ghirlanda et al. 2011. Our model fit to the EGB constrains the so-called `` star formation efficiency of molecular hydrogen", which we found well within existing, much looser observational constraints (Leroy et al. 2008). 

\begin{figure}
\begin{center}
\includegraphics[scale=0.4]{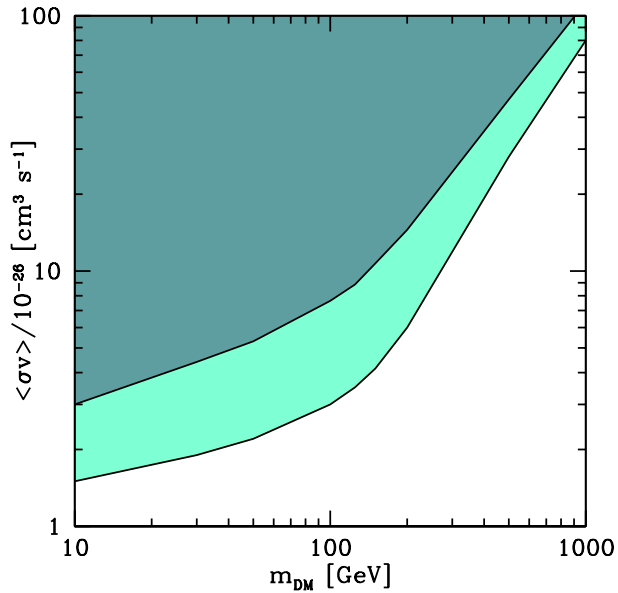}
\end{center}
\caption{Upper limits of the cross section $\langle \sigma v\rangle$ as a function of particle mass $m_{\rm DM}$ for 
annihilating DM.  The lower (upper) curve represents the 1(2)-$\sigma$ limit. See text for details.}
  \label{fig:dm}
\end{figure}

Clearly, the specific best fit values obtained depend upon the details of our model, in terms of star formation rate adopted, models 
for the $\gamma$-ray emission of star-forming galaxies, blazar LF and SED. Nevertheless the overall picture 
appears quite robust, with an important role played by star-forming galaxies, with blazars dominating only at the higher energies probed by {\it Fermi}-LAT. 

Though our model is statistically fully acceptable, 
it is interesting to note that the highest data point of the EGB (see fig. 2 , upper panel) lies above our best fit model. 
In the energy band 50-100 GeV absorption of $\gamma$-rays due to the interaction with the
EBL is significant. Different
theoretical EBL models have been proposed in the last few years 
(e.g., Franceschini, Rodighiero, \& Vaccari 2008; Gilmore et al. 2009; Kneiske \& Dole 2010;  Finke et al. 2010; 
Dom\'inguez et al. 2011), resulting in somewhat different optical depth for photon-photon interaction. 
As already discussed, we follow Finke et al. (2010), and we checked that  even adopting the model of Kneiske \& Dole (2010), which gives the lowest $\gamma$-ray absorption, 
our EGB model still falls short in the 70-100 GeV range. A possible, intriguing explanation is the presence of an extra 
emission from annihilating dark matter (DM) particles (see, e.g., Ullio et al. 2002). 
Recently, Abazajian, Blanchet, \& Harding (2010b) performed a detailed analysis of possible DM candidates in the context of {\it Fermi}-LAT 
EGB.  A full discussion of DM particle contribution to the EGB is beyond the scope of the present {\it Letter}. For illustrative 
discussion, here we adopt a specific annihilating DM model, and compute its contribution to the EGB. 

The $\gamma$-ray background produced by annihilating DM is calculated following Abazajian et al. (2010b) 
and Ando (2005). We found, as an example, that   
a particle of mass $\simeq 0.5$ TeV and cross section $\langle \sigma v \rangle \simeq 5 \times  10^{-26}$ cm$^3$ s$^{-1}$ 
can easly accomodate the last data point. However its presence is not statistically required by the fit, so it is 
fair to consider only upper limits to the DM component. 
Fig.~\ref{fig:dm} shows our results in terms of cross section $\langle \sigma v\rangle$ and particle mass $m_{\rm DM}$. 
The lower (upper) curve is computed by adding the DM background to our EGB model, allowing a 
$\chi^2$ increase of 1 (4) with respect to the best fit, hence representing the 1(2)-$\sigma$ upper limits of 
$\langle \sigma v\rangle$ for a given $m_{\rm DM}$. As an example, assuming  
$\langle \sigma v\rangle = 3\times 10^{-26}$ as required for leaving the
observed relic density of DM (Jungman et al. 1996), we can exclude at 1(2)-$\sigma$ level DM particles with $m_{\rm DM} \lsim 100 (10)$GeV. More massive particles can have a larger cross section, and still be 
compatible with EGB data.  Our limits are consistent with other, more refined, determinations (e.g., Abazajian et al. 
2010b). 

Finally, it is worth asking whether DM particle annihilation signal could substitute the $\gamma$-ray emission from star-forming galaxies in fitting the EGB. We checked that the joint fit to blazar number counts and EGB adopting an alternative 
three component model (i.e., blazars, non-beamed AGNs, and DM)  
is statistically unacceptable ($\chi^2=47.5$ for 28 d.o.f., to be compared to $\chi^{2}=33.3$ for 29 d.o.f. 
as discussed in section 3). We conclude that star-forming galaxies seem to be a necessary, and important, 
component of the EGB, while DM is not strictly required.

		
\section*{acknowledgements}
We have benefited from many informative discussions with M. Ajello, C. D. Dermer, G. Ghisellini, Y. Inoue, and M. Pilia. 
Support to this work was provided by Italian MIUR under PRIN 2007 grant (FH and RS).


\label{lastpage}

\end{document}